\documentclass[floatfix,twocolumn,showpacs,preprintnumbers,amsmath,amssymb,pra,superscriptaddress,longbibliography]{revtex4-1}
\usepackage{color}
\usepackage[usenames,dvipsnames,svgnames,table]{xcolor}
\usepackage[colorlinks=true,linkcolor=blue,urlcolor=blue,citecolor=blue]{hyperref}

\usepackage{mathtools}
\usepackage{graphicx}
\usepackage{dcolumn}
\usepackage{array}
\usepackage{lipsum}
\usepackage{bm}
\usepackage{subfigure}
\usepackage{amssymb}
\usepackage{multirow}
\usepackage{tabularx}
\usepackage{amsmath}
\usepackage{braket}

\newcommand{\br}{\bm{r}}
\newcommand{\bq}{\bm{q}}

\renewcommand{\vec}[1]{\mathbf{#1}}

\begin{document}

\title{Benchmarking Exchange-Correlation Functionals in the Spin-Polarized  Inhomogeneous Electron Gas under Warm Dense Conditions}

\author{Zhandos Moldabekov}
\email{z.moldabekov@hzdr.de}
\affiliation{Center for Advanced Systems Understanding (CASUS), D-02826 G\"orlitz, Germany}
\affiliation{Helmholtz-Zentrum Dresden-Rossendorf (HZDR), D-01328 Dresden, Germany}

\author{Tobias Dornheim}
\affiliation{Center for Advanced Systems Understanding (CASUS), D-02826 G\"orlitz, Germany}
\affiliation{Helmholtz-Zentrum Dresden-Rossendorf (HZDR), D-01328 Dresden, Germany}

\author{Jan Vorberger}
\affiliation{Helmholtz-Zentrum Dresden-Rossendorf (HZDR), D-01328 Dresden, Germany}

\author{Attila Cangi}
\affiliation{Center for Advanced Systems Understanding (CASUS), D-02826 G\"orlitz, Germany}
\affiliation{Helmholtz-Zentrum Dresden-Rossendorf (HZDR), D-01328 Dresden, Germany}

\begin{abstract}
Warm dense matter is a highly active research area both at the frontier and interface of material science and plasma physics. 
We assess the performance of commonly used exchange-correlation (XC) approximation (LDA, PBE, PBEsol, and AM05) in the spin-polarized inhomogeneous electron gas under warm dense conditions based on exact path-integral quantum Monte-Carlo calculations.
This extends our recent analysis on the relevance of inhomogeneities in the spin-unpolarized warm dense electron gas [Z.~Moldabekov \emph{et al.}, J. Chem. Phys. 155, 124116 (2021)].
We demonstrate that the predictive accuracy of these XC functionals deteriorates with (1) a decrease in density (corresponding to an increase in the inter-electronic correlation strength) and (2) an increase of the characteristic wave number of the density perturbation.
We provide recommendations for the applicability of the considered XC functionals at conditions typical for warm dense matter. Furthermore, we hint at future possibilities for constructing more accurate XC functionals under these conditions. 
\end{abstract}

\maketitle

\section{Introduction}
Matter under extreme conditions~\cite{fortov_review, new_pop} is an active area of research both in terms of theory and experiment. Interest in this research area is fueled by its importance for astrophysics \cite{Nettelmann2008,MHVT2008:massive,militzer1,Benuzzi_Mounaix_2014}, controlled fusion \cite{BH16, hu_ICF}, and discovery of novel materials~\cite{Kraus2016,Kraus2017,Lazicki2021}.
In high-energy-density plasma science, extreme conditions are created by various laser- and shock-driven compression techniques~\cite{Fortov2016}. Highly compressed matter at temperatures around the electronic Fermi temperature is referred to as warm dense matter (WDM). Loosely speaking, WDM is a transient state between solids and dense plasmas. As such, understanding the physics in WDM has emerged as a new interdisciplinary challenge for condensed matter and plasma physics~\cite{wdm_book, new_pop}.  

Commonly the Born-Oppenheimer approximation is applied which couples the ion dynamics to the electrons through a potential energy surface in the electronic Schr\"odinger equation. Thus, obtaining an accurate electronic structure is of great significance for WDM modeling.  
Over the years, a wide range of theoretical methods has been developed for dealing with the electronic structure problem at typical WDM parameters. These include path-integral Quantum Monte-Carlo (QMC)~\cite{dornheim_prl,groth_prl17, dornheim_physrep18_0}, restricted QMC~\cite{PhysRevE.103.013203,Militzer_PRL_2012,Brown_PRL_2013}, Kohn-Sham density functional theory (KS-DFT)~\cite{hohenberg-kohn, KS65}, and orbital-free density functional theory \cite{PhysRevLett.113.155006}.

KS-DFT is often the method of choice due to its balance of reasonable accuracy and affordable computational cost. 
The temperature generalization of KS-DFT was originally performed by Mermin~\cite{mermin_65}. Several formal aspects of functional construction at finite temperature have been investigated more recently~\cite{PhysRevLett.107.163001, PhysRevB.84.125118, PhysRevB.93.195132, PhysRevB.93.205140}.
The fundamental ingredient to KS-DFT is the exchange-correlation (XC) functional which needs to be approximated in practice. Hence, the predictive capability of KS-DFT relies on the accuracy of the XC functional. 
Therefore, developing and assessing XC functionals for WDM applications is of pivotal importance.

From a theoretical perspective, the relevance of both electronic correlations and thermal effects strongly impedes our ability to generate accurate data for response properties of electrons in WDM~\cite{wdm_book, new_pop}. 
Obtaining highly accurate, first-principles data for both the static~\cite{dornheim_ML,Dornheim_PRL_2020_ESA,Dornheim_PRL_2020,dornheim_electron_liquid,dornheim_HEDP,dornheim2021density, Dornheim_PRB_2021,Dornheim_PRB_nk_2021} and dynamic response properties~\cite{dornheim_dynamic,dynamic_folgepaper,Dornheim_PRE_2020,Hamann_CPP_2020,Hamann_PRB_2020} of electrons has become possible only very recently by aid of the path-integral QMC method. This, in turn, has allowed us to present the first assessment of commonly used XC functionals in the inhomogeneous electron gas for the relevant WDM parameter range~\cite{moldabekov2021relevance}. 

In the present work we extend our prior analysis by considering the spin-polarized 
inhomogeneous electron gas under WDM conditions.  
To this end, we have performed path-integral QMC calculations of the spin-polarized electron gas under the impact of an external harmonic perturbation~\cite{Dornheim_CPP_2021}.
The generated QMC dataset, primarily the electronic densities, at various wave numbers $q$ and amplitudes $A$ of the external perturbation serves as a reference point to assess the accuracy of several XC functionals. 

Understanding the accuracy of existing XC functionals for spin-polarized electrons is important for a number of reasons~\cite{dornheim2021momentum}.
First of all, the impact of the spin-polarization is crucial for modeling WDM in an external magnetic field. 
Specifically, non-quantizing strong magnetic fields with an amplitude $B\sim 10~{\rm T}...10^4~{\rm T}$ are generated in experiments related to inertial confinement fusion~\cite{Perkins, Appelbe}, where \emph{non-quantizing} means that the characteristic quantum kinetic energy $\sqrt{T_F^2+T^2}$ is dominant over the electron cyclotron energy $\omega_c$~\cite{Haensel} with $T$ denoting the electronic temperature and $T_F$ the Fermi temperature~\cite{quantum_theory,Ott2018}. The latter condition defines the range of non-quantizing magnetic fields as $B/B_0\ll 18.4/r_s^2\sqrt{(T/T_F)^2+1}$ where $B_0\simeq 2.25\times 10^5 {\rm T}$.
Furthermore, electromagnetic pulses applied to WDM can induce spin-polarized states due the effect of the spin-ponderomotive force~\cite{PhysRevLett.105.105004}.
Moreover, the spin-resolved fluid description of quantum plasma dynamics in an external fields requires an XC functional that is capable of describing strongly, non-ideal electrons~\cite{MB2007:dynamics, PhysRevE.91.033111, PhysRevLett.105.105004, Haas_2019, manfredi_fields_05, zhandos_pop18}. Finally, spin-polarized systems are ubiquitous in quantum chemistry and, consequently, benchmarking XC functionals is relevant for modeling hot-electron phenomena~\cite{Brongersma2015}. 

While a wide range of XC functionals is available in the literature, we select the most relevant approximations used for thermal KS-DFT calculations of WDM. We focus on the basic and most commonly used functionals, including the local density approximation (LDA) in the Perdew-Zunger parametrization~\cite{Perdew_LDA} and the generalized gradient approximations PBE~\cite{PBE} and PBEsol~\cite{PBEsol}. Additionally, we consider the Armiento–Mattsson functional (AM05)~\cite{PhysRevB.72.085108} which is a semi-local GGA based on the notion of interpolating between different model systems, namely the uniform electron gas (UEG) and the Airy gas. It was demonstrated for solids~\cite{Mattsson} that AM05 is comparable in its accuracy to the hybrid functionals PBE0~\cite{Perdew96, Adamo2} and HSE06~\cite{Paier}. 
This motivated the use of AM05 for the calculation of the equation of state and electronic structures at WDM parameters~\cite{osti_1055894, Sun, PhysRevB.92.115104}. 

While a number of XC functionals with an explicit temperature dependence has become available recently~\cite{PhysRevB.86.115101, PhysRevLett.112.076403, PhysRevB.88.161108, PhysRevB.88.195103, groth_prl17, Karasiev_PRB_2020}, we purposely do not include them in this benchmark study. We are interested in assessing solely the impact of an inhomogeneous electronic structure due to external perturbations at finite $q$ on the accuracy of ground-state XC functionals. Thus, the temperature dependence is included in our assessment implicitly in terms of a Fermi-Dirac occupation of the KS states. An assessment of explicitly temperature-dependent XC approximations shall be the focus of future studies.

Our paper is organized as follows: we begin with providing the theoretical background and technical details of our calculations in Sec.~\ref{s:theory}; we present our benchmark study in Sec.~\ref{s:results}; finally, we conclude with a summary of our main findings and an outlook on future research in Sec.~\ref{s:end}.

\section{Theory and Simulation Methods}\label{s:theory}
Our assessment of XC functionals is based on the  Hamiltonian of a harmonically perturbed, interacting electron gas~\cite{moroni,moroni2,bowen2,dornheim_pre17,groth_jcp17}
\begin{align}\label{eq:H}
\hat{H} = \hat{H}_{\rm UEG} +  \sum_{k=1}^N \sum_{i=1}^{N_p} 2\, A_i \cos(\hat{\br}_k \cdot {\bq}_i)\ ,
\end{align}
where $\hat{H}_{\rm UEG}$ denotes the standard Hamiltonian of the UEG~\cite{loos,quantum_theory,dornheim_physrep18_0}, $N$ the number of electrons, and $N_p$ the number of harmonic perturbations. 
The strength of the perturbation is controlled by the amplitude $A_i$ and its wavelength by the wave vectors $\vec q_i$.
This Hamiltonian has been used in prior work to investigate fundamental physical properties of both the electron gas and electron liquids, e.g., in terms of the non-linear response~\cite{Dornheim_PRL_2020, Dornheim_PRL_2020_ESA, doi:10.1063/5.0058988, dornheim2021density}, the local field correction (LFC)~\cite{moroni2, moroni, dornheim_pre17, groth_jcp17}, plasmons, and other types of excitations~\cite{moldabekov2021thermal}.
Employing the Hamiltonian in Eq.~(\ref{eq:H}) allows us to asses the performance of different XC functionals in a wide range of conditions ranging from the weakly perturbed UEG where $\delta n/n_0\ll 1$ to the strongly inhomogeneous electron gas with regions of large electronic localization where $\delta n/n_0\gtrsim1$ with $\delta n=n-n_0$.
While any combination of external perturbations is possible, we limit ourselves to an external perturbation with either one harmonic perturbation ($N_p=1$) and a combination of two harmonic perturbations ($N_p=2$). Also, we consider harmonics of the perturbation with the same amplitudes, i.e, $A_i=A$ throughout this work.   

From a physical perspective, we consider typical WDM conditions which correspond to densities $r_s=2$ and $r_s=6$, where $r_s=(4/3 \pi n)^{-1/3}$ is the mean inter-electronic distance in Hartree atomic units.
The temperature is set equal to the Fermi temperature of an unpolarized electron gas~\cite{groth_prl17}. Accordingly, at $r_s=6$ and $r_s=2$ we have $T\simeq 1.4 ~{\rm eV}$ and $T\simeq 12.5 ~{\rm eV}$, respectively. 
These parameters are achieved in solid targets by isochoric heating and laser-induced shock compression experiments~\cite{LANDEN2001465, PhysRevLett.98.065002, PhysRevLett.109.065002}.

The KS-DFT calculations are carried out using the GPAW code~\cite{GPAW1, GPAW2, ase-paper, ase-paper2}. GPAW is a real-space implementation of the projector augmented-wave method which allows us to fix the spin polarization as an external parameter.  
A Monkhorst-Pack~\cite{PhysRevB.13.5188} sampling of the Brillouin zone was used with a \emph{k}-point grid of $8\times 8\times 8$. The calculations are performed using a plane-wave basis where the cutoff energy has been converged to $540~{\rm eV}$.
The number of orbitals is set to $250$ at $r_s=2$ and to $100$ at $r_s=6$. At the considered temperatures, the smallest occupation number at $r_s=2$ is less than $10^{-8}$ and at $r_s=6$ the smallest occupation number is less than $10^{-5}$. It is common practice in thermal KS-DFT calculations to converge the occupation threshold to this value~\cite{doi:10.1063/5.0016538}. Moreover, this criterion was used successfully in our prior study on the spin-unpolarized electron gas where the same Hamiltonian was used~\cite{moldabekov2021relevance}.    

We generate accurate QMC benchmark data by carrying out direct path-integral Monte Carlo calculations \emph{without any nodal restrictions}~\cite{Dornheim_CPP_2021} based on the Hamiltonian in Eq.~(\ref{eq:H}). Therefore, the simulations are extremely costly, but are exact within the given error bars. The compute time is up to $\mathcal{O}\left(10^5\right)$ CPUh for the most challenging parameter sets.
The direct path-integral QMC method is afflicted with the notorious fermion sign problem~\cite{dornheim_sign_problem,troyer}, which leads to an exponential increase in the compute time with increasing the system size $N$ or decreasing the temperature $T$. This problem is further exacerbated in the spin-polarized case, which results in a more pronounced impact of fermionic exchange effects and, therefore, a more severe sign problem.
Yet, we stress that our results are not afflicted with any nodal errors, which have been shown to be considerable in the WDM regime~\cite{Schoof_PRL_2015,lee2020phaseless}. In addition, we note that we have employed a canonical adaption~\cite{mezza} of the worm algorithm by Boninsegni \emph{et al.}~\cite{boninsegni1,boninsegni2}. Finally, we have utilized a straightforward \emph{primitive factorization}~\cite{brualla_JCP_2004,sakkos_JCP_2009} of the density matrix using $P=200$ high-temperature factors. The convergence with $P$ has been checked carefully.

We choose $N=14$ electrons within a simulation cell where its size is defined by $4/3\pi n L^3=1$. 
Accordingly, the smallest wave number of the external perturbation is $q_{\rm min}=2\pi/L\simeq 0.843 q_F$, where $q_F=(3\pi^2 n_0)^{1/3}$ and $n_0$ denotes the mean density when there is no external perturbation.
Note that our calculations do not suffer from finite size effects at the considered parameter range. This has been demonstrated previously in the assessment of the spin-unpolarized electron gas~\cite{moldabekov2021relevance} . Additionally, it was shown in previous QMC studies of the warm dense electron gas~\cite{dornheim_HEDP,dornheim_ML,Dornheim_PRL_2020,dornheim2021density} that the electronic density response is well converged with respect to the number of electrons for as few as $N=14$ electrons.
\begin{figure}
\center
\includegraphics{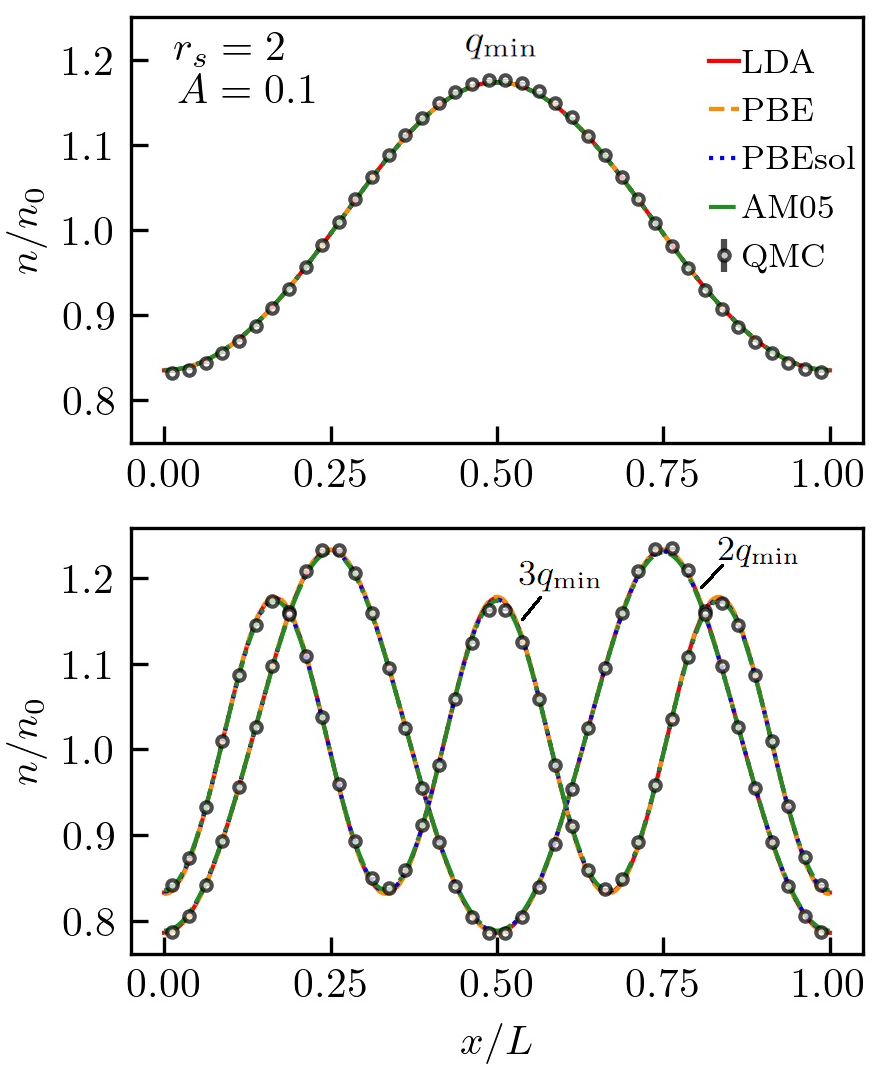}
\caption{ \label{fig:den_rs2_A01} 
Electron density along the perturbation direction for $r_s=2$ and a perturbation with $A=0.1$ where $q_{\rm min}=2\pi/L$ (top), $2q_{\rm min}$ and $3q_{\rm min}$ (bottom).
}
\end{figure} 
\begin{figure}
\center
\includegraphics{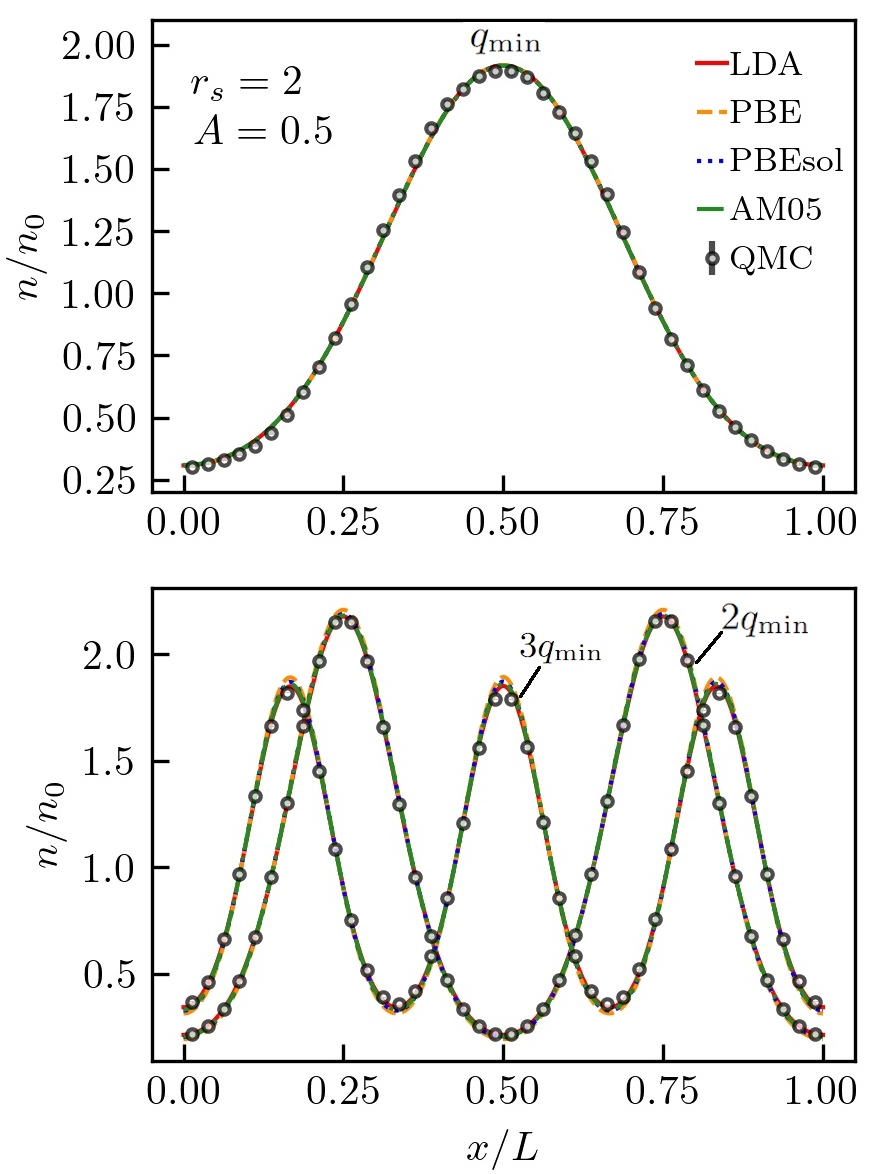}
\caption{ \label{fig:den_rs2_A05} 
Electron density along the perturbation direction for $r_s=2$ and a perturbation with $A=0.5$ where $q_{\rm min}=2\pi/L$ (top), $2q_{\rm min}$ and $3q_{\rm min}$ (bottom).
}
\end{figure} 
\begin{figure}
\center
\includegraphics{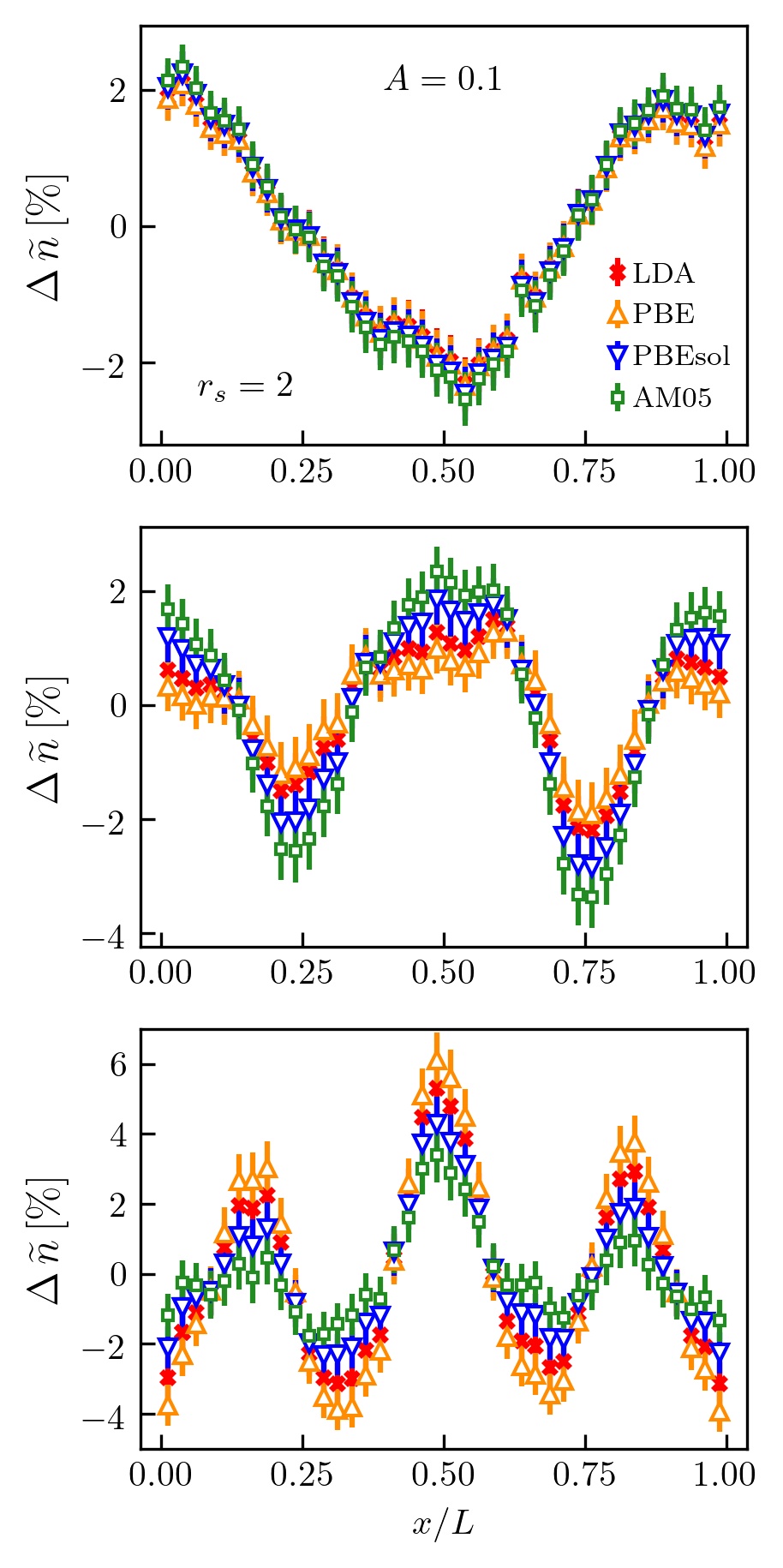}
\caption{ \label{fig:dif_rs2_A01} 
Relative deviation of the electron density at $r_s=2$ in response to an external perturbation with $A=0.1$ and increasing wave numbers $q$. From top to bottom: $q_{\rm min}$, $2q_{\rm min}$, and $3q_{\rm min}$.
}
\end{figure} 
\begin{figure}
\center
\includegraphics{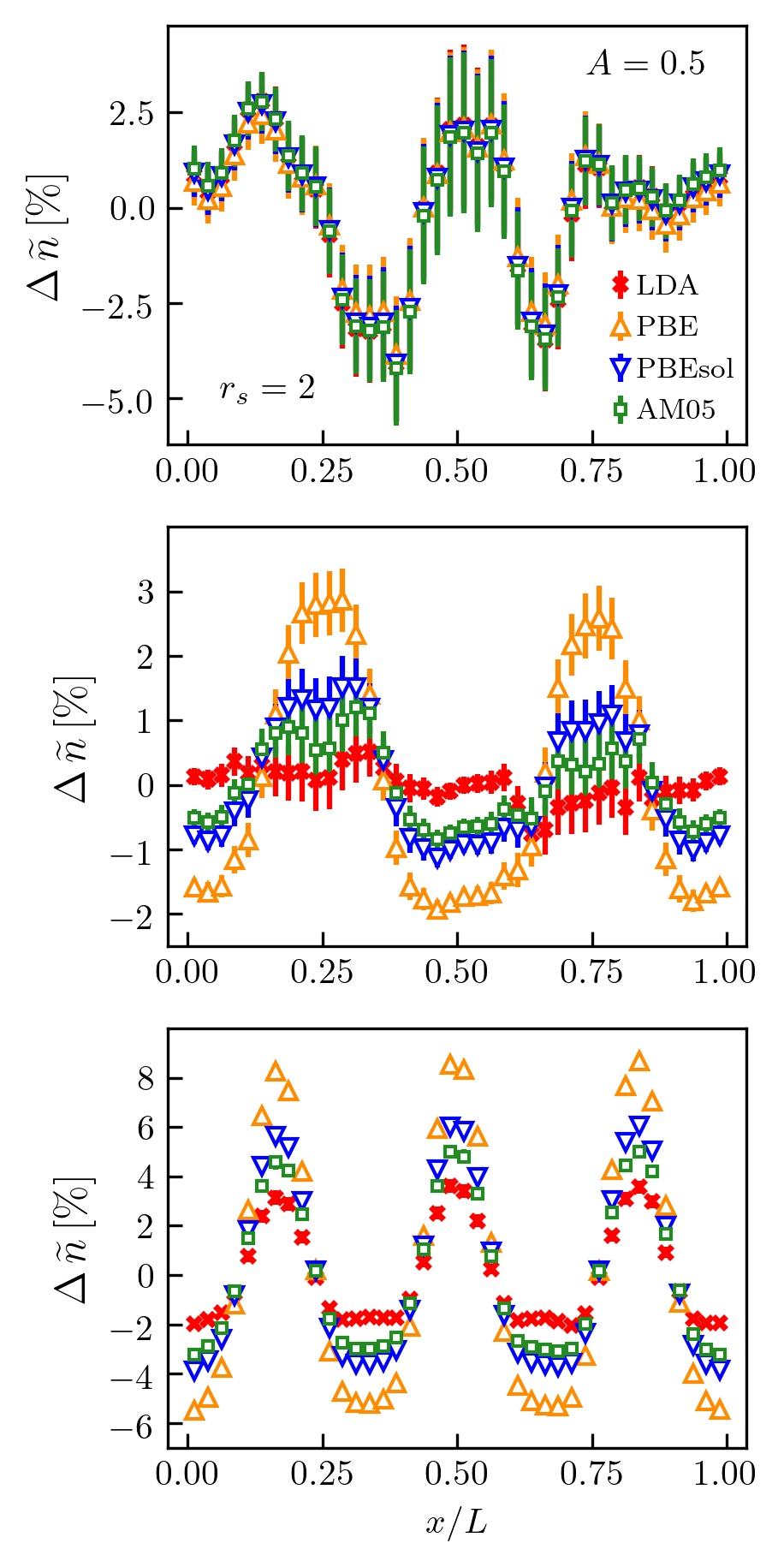}
\caption{ \label{fig:dif_rs2_A05} 
Relative deviation of the electron density at $r_s=2$ in response to an external perturbation with $A=0.5$ and increasing wave numbers $q$. From top to bottom: $q_{\rm min}$, $2q_{\rm min}$, and $3q_{\rm min}$.
}
\end{figure} 
\begin{figure}
\center
\includegraphics{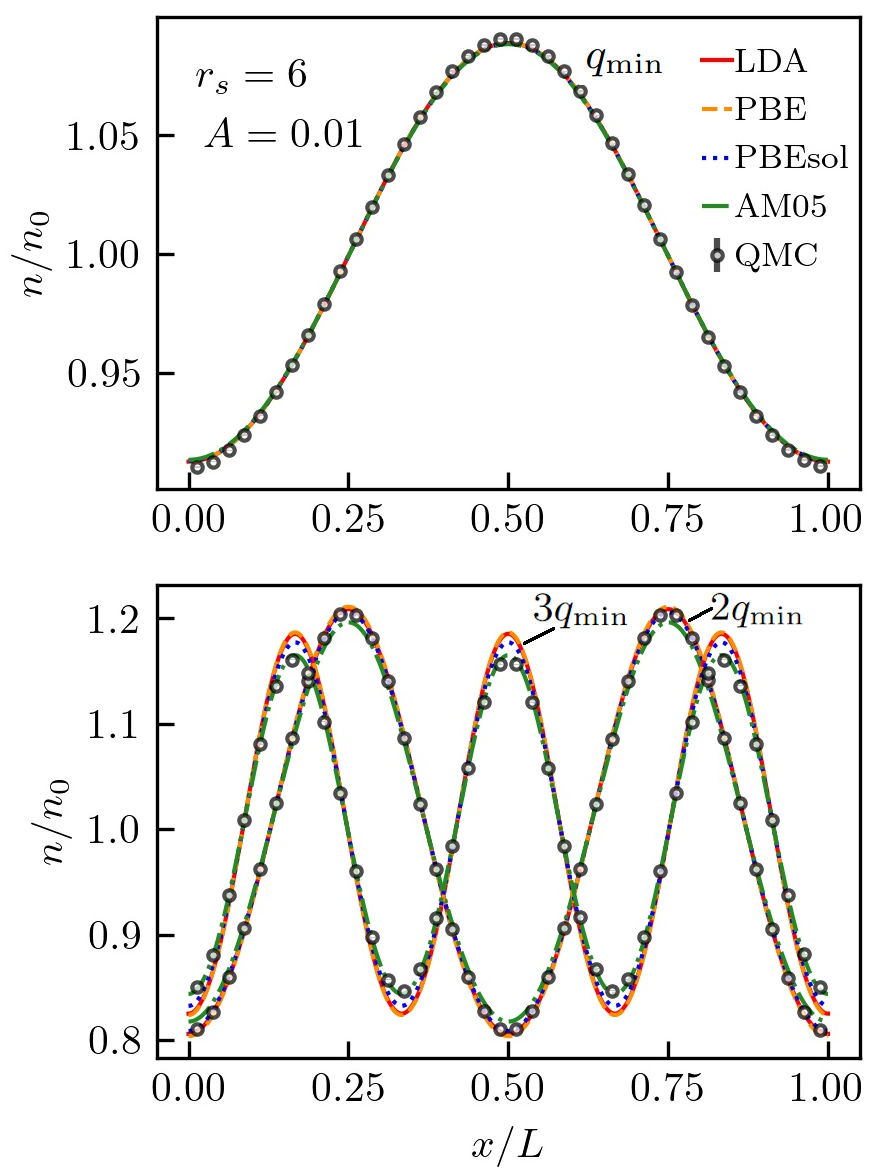}
\caption{ \label{fig:den_rs6_A0_01} 
Electron density along the perturbation direction for $r_s=6$ and a perturbation with $A=0.01$ where $q_{\rm min}=2\pi/L$ (top), $2q_{\rm min}$ and $3q_{\rm min}$ (bottom).
}
\end{figure} 
\begin{figure}
\center
\includegraphics{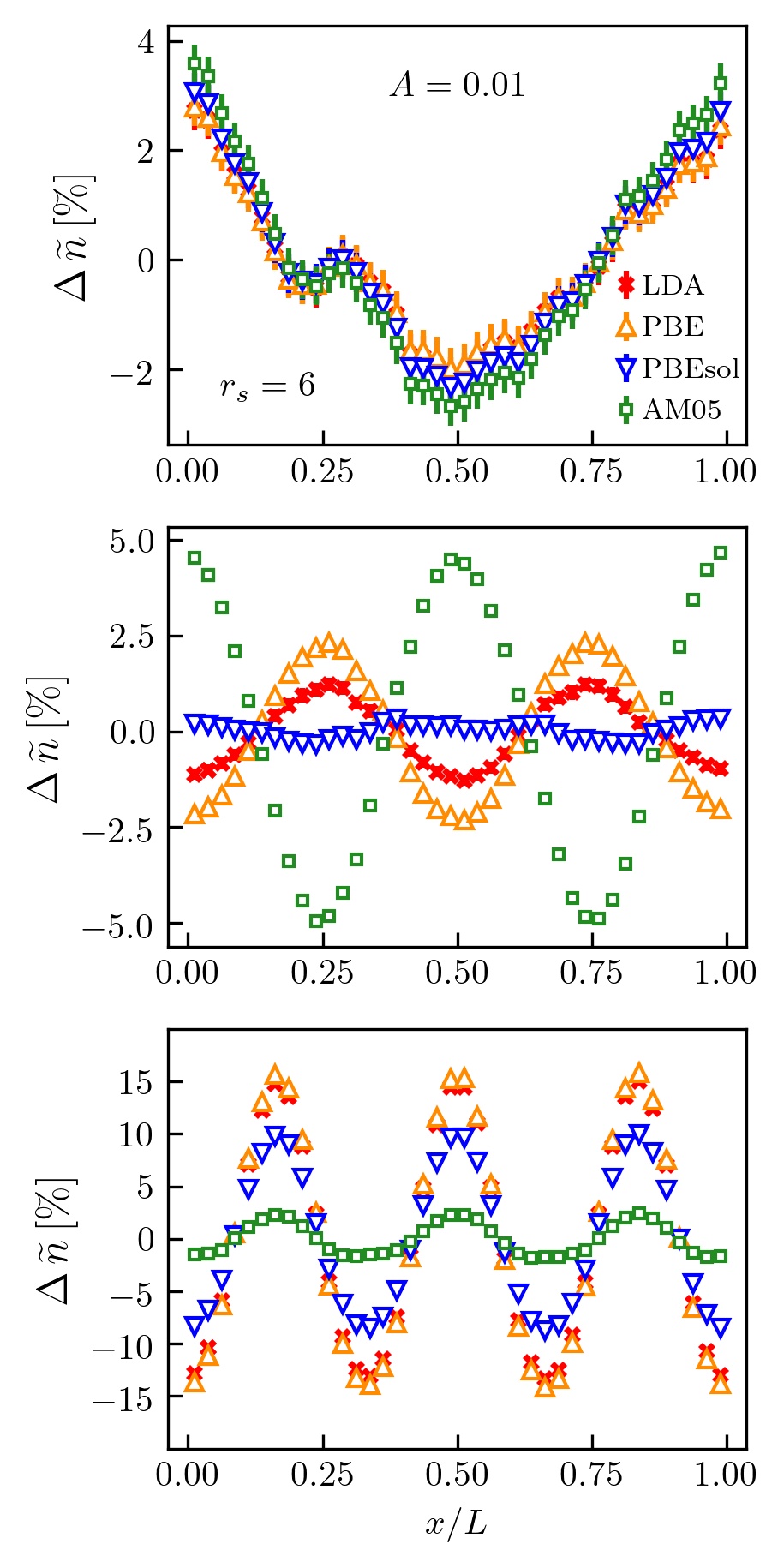}
\caption{ \label{fig:dif_rs6_A0_01} 
Relative deviation of the electron density at $r_s=6$ in response to an external perturbation with $A=0.01$ and increasing wave numbers $q$. From top to bottom: $q_{\rm min}$, $2q_{\rm min}$, and $3q_{\rm min}$.
}
\end{figure}

\section{Results}\label{s:results}
Our assessment of XC functionals centers on the WDM application domain, i.e., partially degenerate plasmas. 
To better quantify these conditions, we consider the ratio of the Coulomb interaction potential between two electrons separated by the mean distance ($1/r_s$) with the Fermi energy ($E_F\sim r_s^{-2}$).
As this is directly proportional to the density parameter $r_s$, we employ $r_s$ as the characteristic coupling parameter of electrons~\cite{Ott2018} for the partially degenerate plasma state considered here.
For ensuring a comprehensive analysis, we perform calculations for both $r_s=2$ and $r_s=6$ which are conditions representative for WDM, where we characterize $r_s=2$ as moderate coupled and $r_s=6$ as strongly coupled regimes.
At $r_s=2$, the kinetic energy of the system is dominant over the XC energy. Contrarily, at $r_s=6$, the XC energy prevails over kinetic energy~\cite{low_density1,low_density2,karasiev_importance}.

\subsection{Moderate coupling, $r_s=2$}
We start with $r_s=2$ and a single harmonic perturbation. This choice of parameters allows us to probe the performance of the XC functionals at different values of the perturbation wave number and amplitude. In principle, any other external perturbation can be represented as a sum of harmonic perturbations. Therefore, the conclusions obtained for the single harmonic perturbation are generally valid and should be transferable. As we show below, this claim is validated by considering the combination of two harmonic perturbations with different wave numbers. 

In Fig. \ref{fig:den_rs2_A01}, we show the total electron density along the direction of the perturbation with an amplitude $A=0.1$ and at three different values of the wave numbers $q_{\rm min}$, $2q_{\rm min}$, and $3q_{\rm min}$.
From the top panel we deduce that the combination of $A=0.1$ and $q_1=q_{\rm min}$ leads to small changes in the electron density $\delta n/n_0<1$. Here, the KS-DFT results are overall in good agreement with the QMC data.
The same is observed for both $q_1=2q_{\rm min}$ and $q_1=3q_{\rm min}$, as shown in the  bottom panel of the same figure. 

Let us next consider a substantially stronger external perturbation. As shown in Fig.~\ref{fig:den_rs2_A01}, we find that an increasing perturbation amplitude ($A=0.5$) results in a stronger density change with $\delta n/n_0\gtrsim1$.
Also here the agreement between KS-DFT and QMC data is quite excellent on the scale of the total density for all considered wave numbers. 
We note that the spin-polarization of the electrons leads to a weaker perturbation at a given amplitude $A$ compared to the spin-unpolarized case considered in Ref.~\cite{moldabekov2021relevance}. The physics behind this observation is discussed in detail in the Appendix. 

\begin{table}[]
\caption{The performance of common XC functionals in terms of the relative deviation of the density $\Delta \widetilde n ~[\%]$. At fixed density $r_s=2$ a single harmonic perturbation with perturbation amplitudes $A=0.1$ and $A=0.5$ and wave numbers $q_{\rm min}\leq q\leq 3 q_{\rm min}$ is considered, where $q_{\rm min}=0.843~q_F$. The largest absolute values of the deviation are listed in this table. Note that at $A=0.1$ ($A=0.5$), the uncertainty is $\pm 0.39\%$ ($\pm1.5\%$) at $q_1=q_{\rm min}$, $\pm 0.55\%$ ($\pm 0.5\%$) at $q_1=2q_{\rm min}$, and $\pm 0.78\%$ ($\pm0.3\%$) at $q_1=3q_{\rm min}$.}
\vspace{0.5cm}
\label{Table_rs2}
\begin{tabular}{ c m{1em} ccc m{1em} ccc}
\hline   
\hline   
\multirow{2}{*}{} & & \multicolumn{3}{c}{$\mathbf{A=0.1}$} & & \multicolumn{3}{c}{$\mathbf{A=0.5}$} \\
\hline
& & $\bf q_{\rm min}$   & $\bf 2 q_{\rm min}$ & $\bf  3 q_{\rm min}$ 
& & $\bf q_{\rm min}$   & $\bf 2 q_{\rm min}$ & $\bf 3 q_{\rm min}$\\ 
 \hline 
{\bf LDA}    & &  2.32 &  2.20 & 5.31& & 4.10 & 0.76 & 3.61\\ 
\hline
{\bf PBE}    & & 2.34 &  1.90 & 6.12 & & 3.85 & 2.86 &  8.68 \\ 
\hline
{\bf PBEsol} & & 2.46 &  2.85 &  4.28 & &  4.10 & 1.51 &  6.05 \\ 
\hline
{\bf AM05}   & & 2.54 &  3.36 & 3.39 & & 4.20 & 1.20 & 5.01 \\ 
\hline
\hline
\end{tabular}
\end{table}

To gain deeper insight, we consider the density perturbation $\Delta \widetilde n$ which is the physically more important quantity here. It defines the (linear) static response function in the case of a weak external field since $\delta n=2A \chi(\vec q) \cos(\vec q\cdot \vec r)$. Therefore, we further quantify differences in the density perturbation by considering  the relative density deviation between the KS-DFT data and the reference QMC data defined as
\begin{equation}\label{eq:Dn}
    \Delta \widetilde n ~[\%]=\frac{\delta n_{\rm DFT}-\delta n_{\rm QMC}}{{\rm max}\{\delta n\}} \times 100,
\end{equation}
where ${\rm max}\{\delta n\}$ is the maximum deviation of the QMC data from the mean density. This quantity provides a meaningful point of reference to gauge the quality of the DFT data.

In Figs. \ref{fig:dif_rs2_A01} and \ref{fig:dif_rs2_A05} we illustrate the relative density deviation $\Delta \widetilde n$ for $r_s=2$ at perturbation amplitudes $A=0.1$ and $A=0.5$ and different values of the perturbing wave number. The maximum values of the relative deviations are also listed in Table \ref{Table_rs2}.
At $q_1=q_{\rm min}$  and $q_1=2q_{\rm min}$, the maximum deviation of the KS-DFT data from the QMC data is about $2\%$ and all considered XC functionals provide a similar quality within the given statistical uncertainty of the QMC data. When we increase the perturbing wave number to $3q_{\rm min}$, we observe larger errors in the KS-DFT data, where the calculations based on the PBE functional exhibit the largest errors of about $6\%$. In contrast to that, the AM05 functional performs well with a maximum deviation of about $3\%$. The calculations based on LDA and PBEsol functionals yield a comparable accuracy in $\Delta \widetilde n$ with maximum deviations of $5\%$ and $4\%$, respectively.   
Increasing the perturbation strength to $A=0.5$, as shown in Fig.~\ref{fig:dif_rs2_A05}, generally leads to less accurate KS-DFT results. The PBE performs worst for the two largest considered wave numbers. At $A=0.5$ and $q_1=3q_{\rm min}$, the error in PBE is close to $9\%$, while the LDA yields an error of about $4\%$. 

Within the statistical uncertainty of the QMC data, the overall quality of the KS-DFT calculations  based on LDA, PBE, and AM05 at $A=0.5$ is similar to that at $A=0.1$. 
Nevertheless, we can single out LDA as the XC functional providing the best quality on average for $q_1=q_{\rm min}$, $q_1=2q_{\rm min}$, and $q_1=3q_{\rm min}$. Additionally, in both cases $\delta n/n_0<1$ and $\delta n/n_0\gtrsim1$, we conclude that all considered XC functionals provide an accurate description of the total density at perturbation wave numbers $q\leq 2q_{\rm min}\simeq 1.7 q_F$. 
At larger wave numbers $3q_{\rm min}$ the considered XC functionals start becoming less accurate.

\subsection{Strong coupling, $r_s=6$}
Next, we investigate the strongly coupled regime at $r_s=6$. We again consider a single harmonic perturbation with different amplitudes and wave numbers.
The strongly coupled regime turns out to be more challenging for the XC functionals under examination.

We begin with considering $\delta n/n_0\ll 1$ which is achieved by a weak perturbation with $A=0.01$. The total electron density along the perturbation direction is shown in Fig.~\ref{fig:den_rs6_A0_01} for three different values of the perturbation wave number,  $q_{\rm min}$, $2q_{\rm min}$, and $3q_{\rm min}$.  
While at small wave numbers, such as $q_1=q_{\rm min}$ in the top panel, all XC functionals perform well, significant deviations from the QMC reference data become evident for most XC functionals at larger wave numbers, e.g., at $2q_{\rm min}$  and $3q_{\rm min}$ (bottom panel).

In Fig.~\ref{fig:dif_rs6_A0_01}, we further examine the disagreement between the KS-DFT and QMC calculations using the relative density deviation.
We infer that at $q_1=q_{\rm min}$ all considered XC functionals have a maximum deviation of about $2\%$ in the central region around the maximum of the density and deviate up to about $3\%$ around the density minima.  At $q_1=2q_{\rm min}$, the AM05 functional has an error abour $5\%$ around the density minima and maxima. 
Contrarily, at $q_1=2q_{\rm min}$ other XC functionals provide a rather accurate description, the PBE data yielding a maximum error of about $2\%$ and both the LDA and PBEsol being virtually exact.
When the perturbing wave number increases to $3q_{\rm min}$, the LDA, PBE, and PBEsol functionals suffer from large errors as shown in the bottom panel of Fig.~\ref{fig:dif_rs6_A0_01}. Specifically, LDA and PBE have an error of about $15\%$ and PBEsol of up to $10\%$.
In contrast to that and to the case of small wave numbers, the AM05 functional provides relatively accurate results with a maximum error of about $2.5\%$ at large wave numbers. 

The high accuracy of the KS-DFT calculations based on the considered XC functionals in the case of a weak perturbation is due to connection of the XC energy density with the long wavelength of XC kernel of the static response function via the compressibilty sum-rule \cite{PhysRevB.88.115123}. For the same reason, the accuracy of the KS-DFT results becomes worse with an increasing wave number of the perturbation. We discuss this point in more detail in Sec.~\ref{s:end}.

It is noteworthy that the AM05 functionals performs better for perturbations at large wavenumbers (compare the bottom panel of Fig.~\ref{fig:dif_rs6_A0_01} with the top and middle). This can be understood heuristically by the fact that AM05 is based on an two electronic reference systems, namely the UEG and the Airy electron gas \cite{PhysRevLett.81.3487}. It interpolates between the slowly varying limit in jellium bulk, $q\to0$, and the limit far from the jellium surface with a characteristic wave number $q/q_F\gg 1$ representing the density changes. The latter is the reason why the AM05 functional performs better at $3q_{\rm min}$. Note that this assessment is valid at $r_s=2$ and $r_s=6$ if the perturbation is weak $\delta n/n_0<1$, but not when the density perturbation is strong $\delta n/n_0>1$. We extend this discussion towards the end in Sec.~\ref{s:end}.

\begin{figure}
\center
\includegraphics{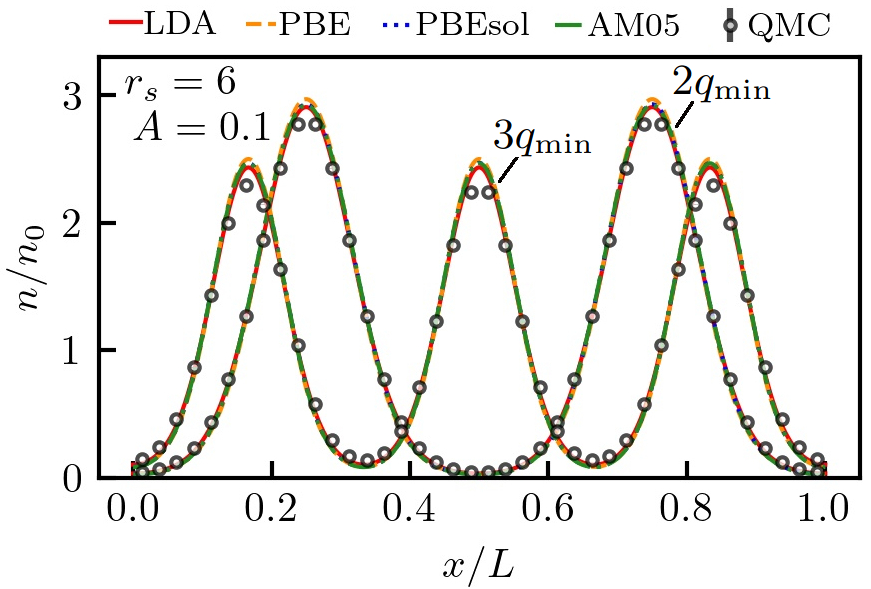}
\caption{ \label{fig:den_rs6_A0_1} 
Electron density along the perturbation direction for $r_s=6$ and a perturbation with $A=0.1$ at $2q_{\rm min}$ and $3q_{\rm min}$, $q_{\rm min}=2\pi/L$.
}
\end{figure} 

\begin{figure}
\center
\includegraphics{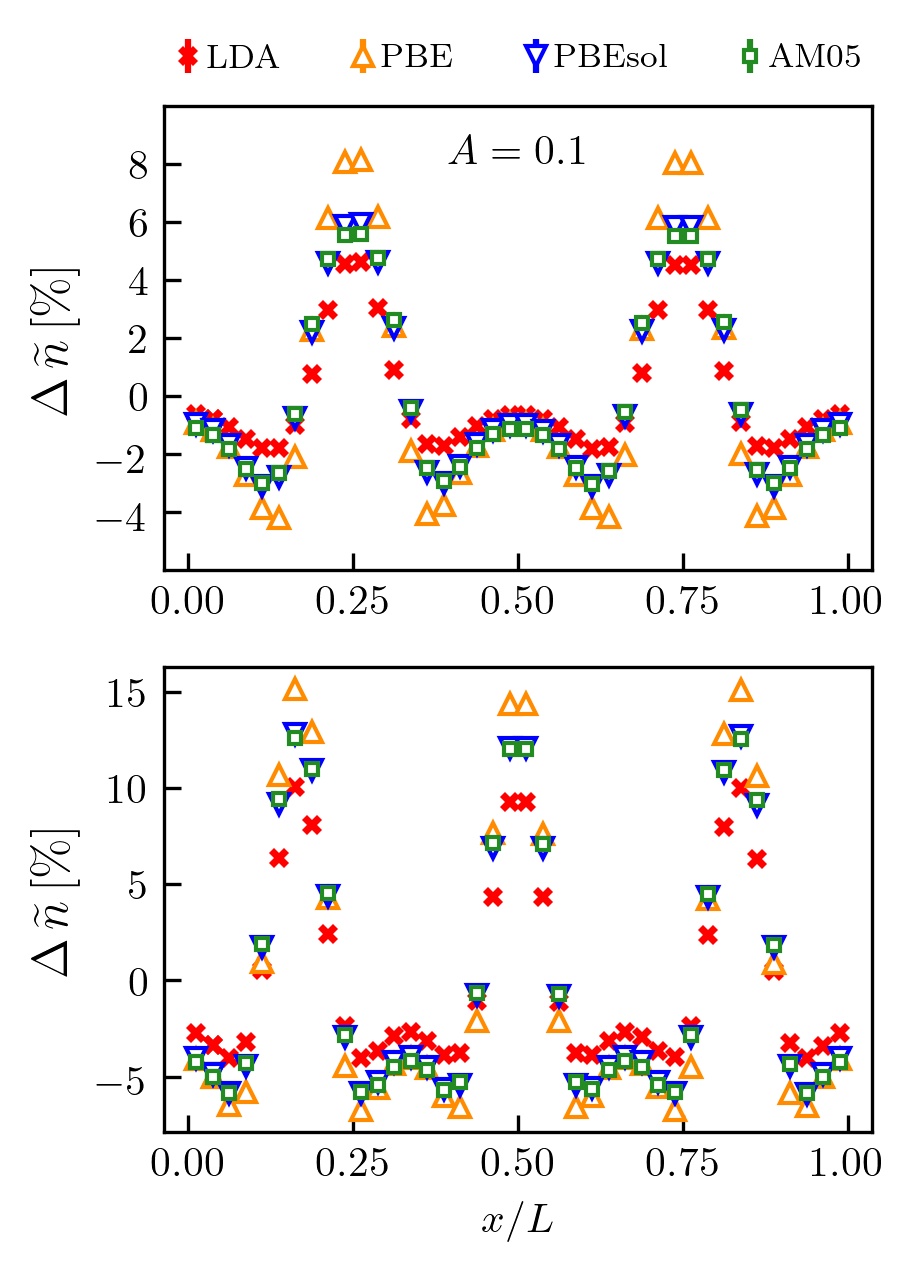}
\caption{ \label{fig:dif_rs6_A0_1} 
Relative deviation of the electron density at $r_s=6$ in response to an external perturbation with $A=0.1$ and increasing wave numbers $q$. From top to bottom: $2q_{\rm min}$ and $3q_{\rm min}$.
}
\end{figure}

Next, we consider strong perturbations $A=0.1$ which cause strong density changes $\delta n/n_0>1$ in Fig.~\ref{fig:den_rs6_A0_1}. It illustrates the total density at the perturbation wave numbers $2q_{\rm min}$ and $3q_{\rm min}$. 
Similar to weak perturbations, the KS-DFT results differ significantly from the QMC data.
We analyze these differences further by computing the the relative density deviations shown in Fig.~\ref{fig:dif_rs6_A0_1}.
Here, the PBE functional performs worst with maximum errors of about $8\%$ and $15\%$ maximum deviations at $2q_{\rm min}$ and $3q_{\rm min}$, respectively. The AM05 and PBEsol functionals exhibit both comparable errors of up to $6\%$ at $2q_{\rm min}$ and about $13\%$ at $3q_{\rm min}$.  The LDA performs best with a maximum error of about $4\%$ at $2q_{\rm min}$, but still with large errors of about $10\%$ at $3q_{\rm min}$.

At the smallest considered wave number of the perturbation, $q_{\rm min}$, the KS-DFT data is relatively accurate with a maximum error of about $3\%$ and $4\%$. The maximum values of the relative density deviation are listed in Table~\ref{Table_rs6}.

We conclude our assessment of the strong-coupling regime by summarizing that LDA, PBE, and PBEsol provide an accurate description of the total density for relatively small wave numbers $q\leq 2q_{\rm min}\simeq 1.7 q_F$ and small density changes $\delta n/n_0<1$. We point out that AM05 does not show such a clear trend with a change in the perturbing wave number when $\delta n/n_0<1$. We provide a heuristic explanation in the Sec.~\ref{s:end}. 
However, when the perturbation is strong ($\delta n/n_0>1$), all considered XC functionals yield significant errors with respects to the QMC reference data at both $q= 2q_{\rm min}\simeq 1.7 q_F$ and $q=3q_{\rm min}\simeq 2.53 q_F$.

\begin{table}[]
\caption{The performance of common XC functionals in terms of the relative deviation in the density $\Delta \widetilde n ~[\%]$. A single harmonic perturbation at a fixed density $r_s=6$, perturbation amplitudes $A=0.01$ and $A=0.1$, and wave numbers $q_{\rm min}\leq q\leq 3 q_{\rm min}$ is considered, where $q_{\rm min}=0.843~q_F$. The largest absolute values of the error are listed. Note that at $A=0.1$ ($A=0.5$), the uncertainty in the QMC data is $\pm 0.35\%$ ($\pm0.25\%$) at $q_1=q_{\rm min}$, $\pm 0.18\%$ ($\pm 0.13\%$) at $q_1=2q_{\rm min}$, and $\pm 0.25\%$ ($\pm0.11\%$) at $q_1=3q_{\rm min}$.}
\vspace{0.5cm}
\label{Table_rs6}
\begin{tabular}{ c m{1em} ccc m{1em} ccc}
\hline   
\hline   
\multirow{2}{*}{} & & \multicolumn{3}{c}{$\mathbf{A=0.01}$} & & \multicolumn{3}{c}{$\mathbf{A=0.1}$} \\
\hline
& & $\bf q_{\rm min}$   & $\bf 2 q_{\rm min}$ & $\bf  3 q_{\rm min}$ 
& & $\bf q_{\rm min}$   & $\bf 2 q_{\rm min}$ & $\bf 3 q_{\rm min}$\\ 
 \hline 
{\bf LDA}    & &  2.72 &  1.28 & 15.0 & & 3.51 & 4.6 & 10.05\\ 
\hline
{\bf PBE}    & & 2.79 &  2.31 & 15.83 & & 3.56 & 8.11 &  15.20 \\ 
\hline
{\bf PBEsol} & & 3.06 &  0.34 &  10.0 & &  3.44 & 5.91 &   12.80 \\ 
\hline
{\bf AM05}   & & 3.58 &  4.95 & 2.42 & & 2.92 & 5.59 &  12.61\\ 
\hline
\hline
\end{tabular}
\end{table}

\subsection{Combined effect of two harmonic perturbations}\label{app:two_perturbations}

\begin{figure}
\center
\includegraphics{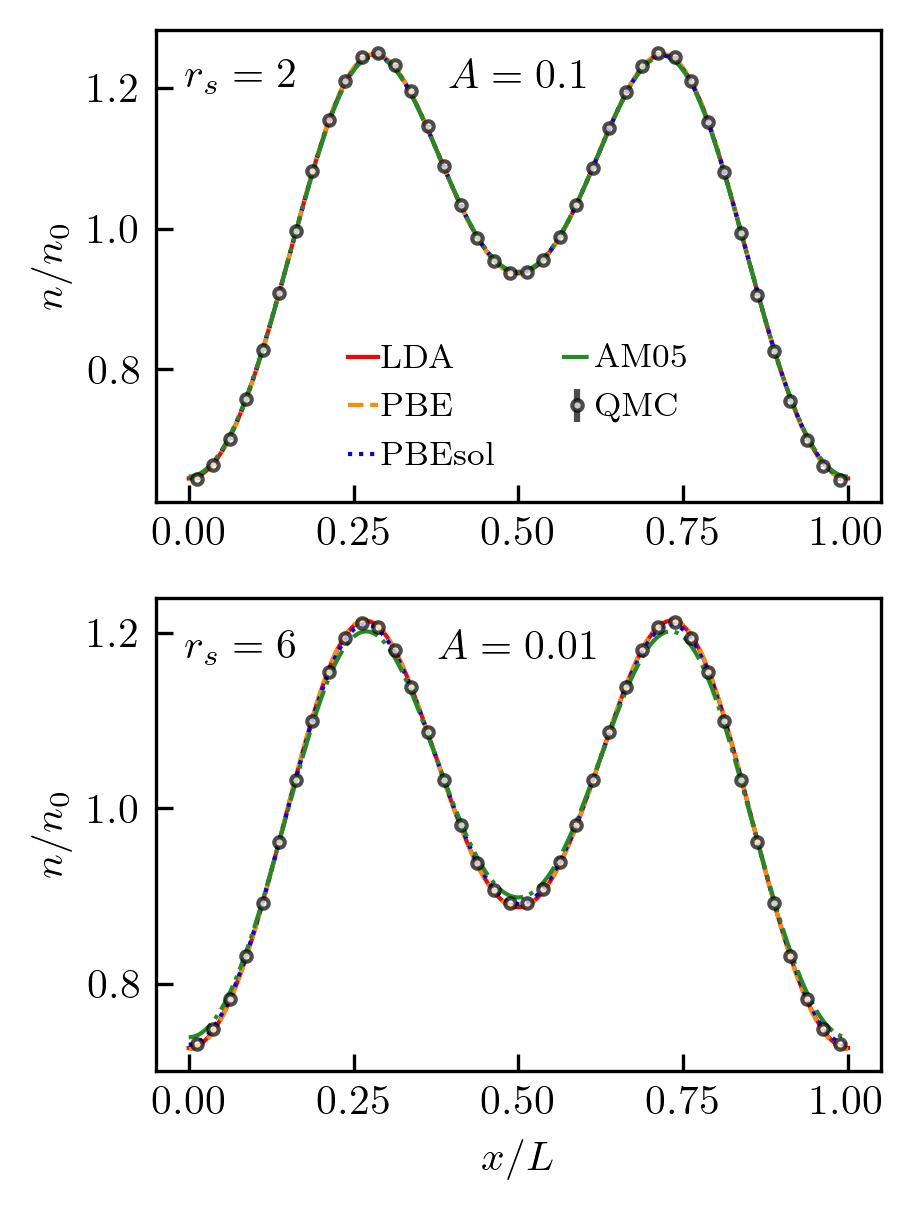}
\caption{ \label{fig:den_double} 
Electron density along the perturbation direction for a double harmonic perturbation with wave numbers $q_1=q_{\rm min}$ and $q_2=2q_{\rm min}$ and amplitudes $A=0.1$ and $A=0.01$ for moderate coupling $r_s=2$ (top) and strong coupling $r_s=6$ (bottom).
}
\end{figure} 

\begin{figure}
\center
\includegraphics{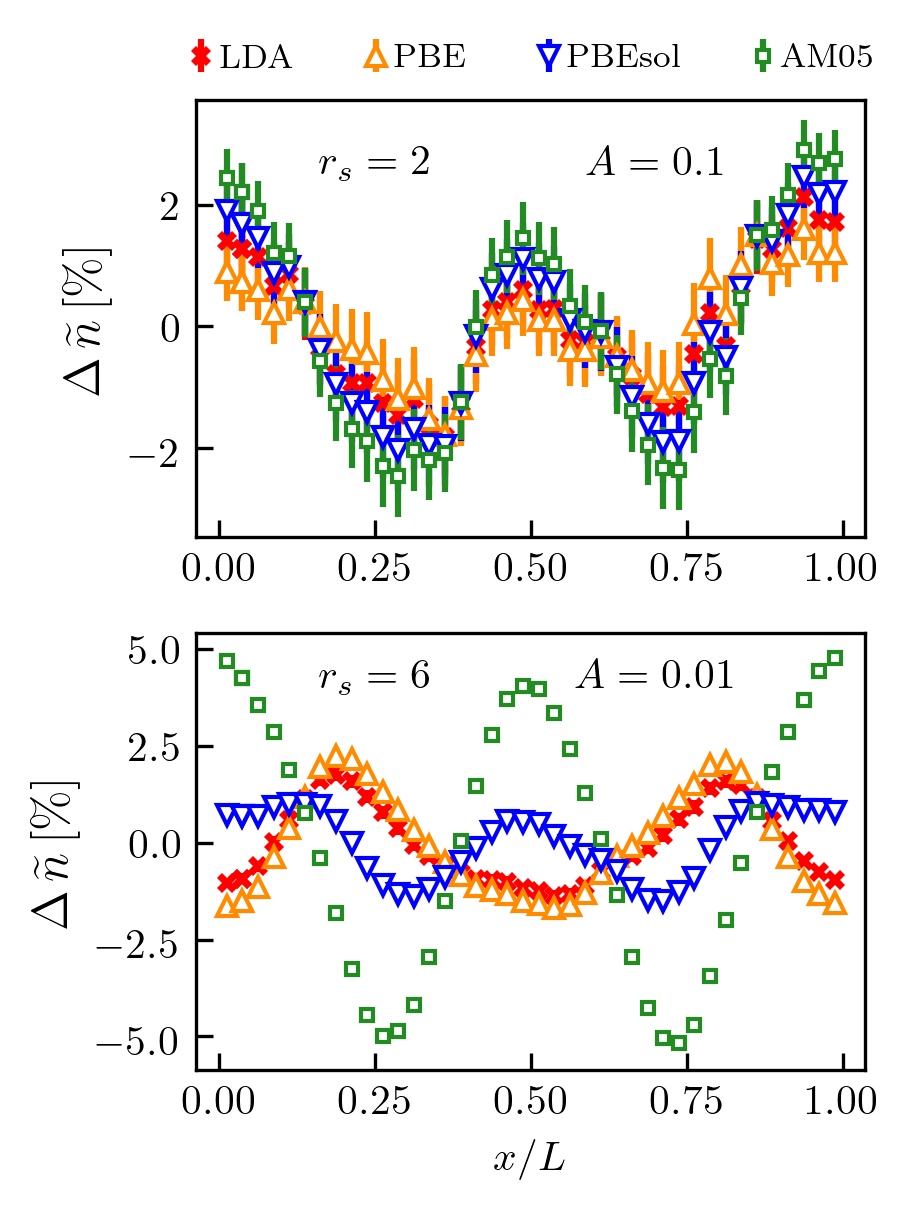}
\caption{ \label{fig:dif_double} 
Relative deviation of the electron density in response to an external double harmonic perturbation with wave numbers $q_1=q_{\rm min}$ and $q_2=2q_{\rm min}$ and amplitudes $A=0.1$ and $A=0.01$ for moderate coupling $r_s=2$ (top) and strong coupling $r_s=6$ (bottom).
}
\end{figure}

In general, any perturbation can be represented as a linear combination of harmonic perturbations. To provide supporting evidence that our analysis is not an artefact of using a single harmonic perturbation, we consider the combination of two harmonic perturbations with the wave numbers $q_1=q_{\rm min}$ and $q_2=2q_{\rm min}$. Note that we choose the same amplitudes, namely $A=0.1$ in the moderately coupled regime ($r_s=2$) and $A=0.01$ in the strongly coupled regime ($r_s=6$). 
These correspond to weakly perturbed states for which the analysis based on the single harmonic perturbation showed failure of AM05 at  $r_s=6$ and $2q_{\rm min}$ (see middle panel in Fig. ~\ref{fig:dif_rs6_A0_01}). Now, we can test whether the same holds when the perturbation is the combination of harmonics with different wave numbers. 

The resulting electron density and the relative density deviation are shown in Figs.~\ref{fig:den_double} and Fig.~\ref{fig:dif_double}.
From these we infer that LDA, PBE, and PBEsol provide an accurate description of the total density for both $r_s=2$ and $r_s=6$ when the perturbation is weak.
The same is valid for AM05 at $r_s=2$. However, AM05 becomes less accurate for strong coupling ($r_s=6$).
We note that maximum errors in the relative density deviation follow a similar trend as observed for a single harmonic perturbation with the same values of the perturbation amplitudes (cf. Figs~\ref{fig:dif_rs2_A01} and \ref{fig:dif_rs6_A0_01}). 
Therefore, we confirm that our conclusions based on a single harmonic perturbation are generally applicable.

\section{Conclusions and Outlook}\label{s:end}

We have benchmarked the accuracy of KS-DFT calculations based on the LDA, PBE, PBEsol, and AM05 XC functionals against exact QMC data for a spin-polarized, partially degenerate, inhomogeneous electron gas in the regime of both weak and strong perturbations. Furthermore, we have considered both moderate and strong coupling regimes which are relevant for current WDM applications. 
We stress that a comprehensive assessment of the vast amount of the available XC functionals has not been attempted in this work. Instead, we focused on the LDA functional and it is generalizations, which are most often used for WDM simulations. 
We therefore provide access to our QMC data to facilitate assessments of any existing or newly developed XC functional by the scientific community~\cite{data}.
With the present work devoted to the spin-polarized case and our earlier work to the spin-unpolarized case~\cite{moldabekov2021relevance}, we conclude assessing common, ground-state XC functionals within the harmonically perturbed, warm dense electron gas.

Our analysis has revealed (1) a parameter range where the KS-DFT method can be used to obtain QMC-level accuracy and (2) conditions when the considered XC functionals become inaccurate and eventually fail to yield an accurate electron density. 

We summarize the main conclusions regarding the applicability of the considered XC functionals as follows: 
\begin{itemize}
    \item For moderate coupling ($r_s=2$) and perturbing wave numbers $q\lesssim 2.53 q_F$, LDA, PBEsol, and AM05 provide an accurate electron density for both weak ($\delta n/n_0<1$) and strong ($\delta n/n_0>1$) density perturbations. 
    \item For strong coupling ($r_s=6$) and weak density perturbations ($\delta n/n_0<1$), LDA, PBE, and PBEsol yield an accurate electron density at $q\leq 2q_{\rm min}\simeq 1.7 q_F$. AM05, however, becomes less reliable at $q= 2q_{\rm min}\simeq 1.7 q_F$. 
    At larger perturbing wave numbers ($q \simeq  2.5q_{F}$), LDA, PBE, and PBEsol are unreliable, while AM05 provides significantly more accurate description than LDA, PBE, and PBEsol.    
    \item  For strong coupling ($r_s=6$) and strong density perturbations ($\delta n/n_0>1$), all considered XC functionals provide an accurate electron density at small perturbing wave numbers $q<q_F$. However, all of them become unreliable for larger wave numbers, $q>q_F$.  
\end{itemize}

Some computational aspects regarding the choice of XC functional are discussed in Appendix~\ref{app:computational_scaling}. While this discussion is rather technical, we believe it can be helpful for users of KS-DFT to save compute time and resources.

Apart from delivering useful benchmarks, our assessment advances our understanding on the expected accuracy of KS-DFT when used for spin-polarized simulations at parameters relevant for WDM applications. 
Clearly, the considered XC functionals struggle to provide accurate results for perturbations with large wave numbers. These conditions are, however, relevant to contemporary WDM experiments where the impact of the external driving fields and of shock formation causes such perturbations in the electronic structure. 

We can rationalize the observed errors at large wave numbers by considering the static linear response function. It is defined as $\chi(q)^{-1}=\chi_0(q)^{-1}-\left[4\pi/q^2 + f_{\rm xc}(q)\right]$, where $\chi_0$ denotes the response function of the UEG and $f_{\rm xc}(q)$ the XC kernel. Using the LDA is equivalent to working within the long wavelength limit of the XC kernel $f_{\rm xc}(q\ll0)$, which is connected to the XC energy via compressibility sum rule~\cite{PhysRevB.88.115123}. The same applies to PBE and PBEsol, for which exchange and correlation gradient corrections are designed to cancel out each other in the limit of the UEG~\cite{PhysRevB.77.245107}. 
Therefore, our analysis points out a significant deficiency of common XC functionals in terms of the wave number $q$. When perturbations at values larger than the Fermi wave number are present, common XC functionals will not be reliable. 

An interesting trend is observed for AM05 at $\delta n/n_0<1$. The accuracy of AM05 first drops with increasing wave number from $q_{\rm min}\simeq 0.84 q_F$ to $2q_{\rm min}$. However, with a further increase in the wave number to $3q_{\rm min}$, its accuracy improves. 
We can shed light into this behavior by looking more closely how AM05 is constructed. The exchange energy in AM05 is an interpolation of two subsystems: the LDA exchange energy and the exchange energy of the Airy gas.
Similarly, the correlation energy of AM05 is interpolated between LDA correlation and a scaled LDA correlation such that it reproduces the jellium surface energy. 
As an integral part of the AM05 functional, the Airy gas is a model system which describes the electrons on surfaces. It is characterized by a decaying electron density with increasing distance from the surface. The connection between the UEG and Airy gas is established using the characteristic wave number of the density variation $q^{\prime}\sim |\nabla n|/n$. Essentially, AM05 reproduces the strongly inhomogeneous Airy gas at large wave numbers $s=q^{\prime}/(2q_F)>1$. Therefore, in our assessment, AM05 performs better than LDA, PBE, and PBEsol, for weak density perturbations $\delta n/n_0<1$ with $3q_{\rm min}>2q_F$. This is no longer the case when the perturbation is strong $\delta n/n_0>1$. This might be related to the fact that only the exchange energy of the Airy gas was used in the construction of AM05, while its correlation energy is based on a scaled LDA due to the lack of exact QMC data for the Airy gas. 
Therefore, this result hints at the need for accurate QMC data for the Airy gas. 
Furthermore, our findings reveal that the interpolation scheme used for AM05 struggles with density perturbations at intermediate wave numbers $q_F<q<2q_F$ when $\delta n/n_0<1$. At $\delta n/n_0>1$, the AM05 is less accurate than LDA. 
We note that these conclusions are valid also for unpolarized electrons. An in-depth analysis of the AM05 parametrization shall be presented elsewhere.  

Our presented findings suggest that interpolating between reference systems in the fashion of the AM05 functional might be a promising route to XC functionals accurate across a large range of wave numbers needed for WDM applications.

Finally, we note that our conclusions are relevant for further developing the non-linear response theory of the electron gas and liquids at WDM conditions and beyond~\cite{Dornheim_PRL_2020, doi:10.1063/5.0058988, dornheim2021density}. 
Indeed, various available QMC methods are restricted with respect to the parameter range due to the infamous fermion sign problem~\cite{dornheim_physrep18_0, dornheim_sign_problem,dornheim2021fermion,troyer,Loh_sign_problem}. 
KS-DFT  allows us to study non-linear response phenomena at temperatures $T<T_F$ and densities $2\lesssim r_s\lesssim 6$ that are difficult to reach with QMC methods.


\section*{Acknowledgments}
This work was funded by the Center for Advanced Systems Understanding (CASUS) which is financed by the German Federal Ministry of Education and Research (BMBF) and by the Saxon Ministry for Science, Art, and Tourism (SMWK) with tax funds on the basis of the budget approved by the Saxon State Parliament. We gratefully acknowledge computation time at the Norddeutscher Verbund f\"ur Hoch- und H\"ochstleistungsrechnen (HLRN) under grant shp00026, and on the Bull Cluster at the Center for Information Services and High Performance Computing (ZIH) at Technische Universit\"at Dresden.

\appendix

\section{Spin-polarized versus spin-unpolarized electron gas}\label{app:comparison_spinpol} 

\begin{figure}
\center
\includegraphics{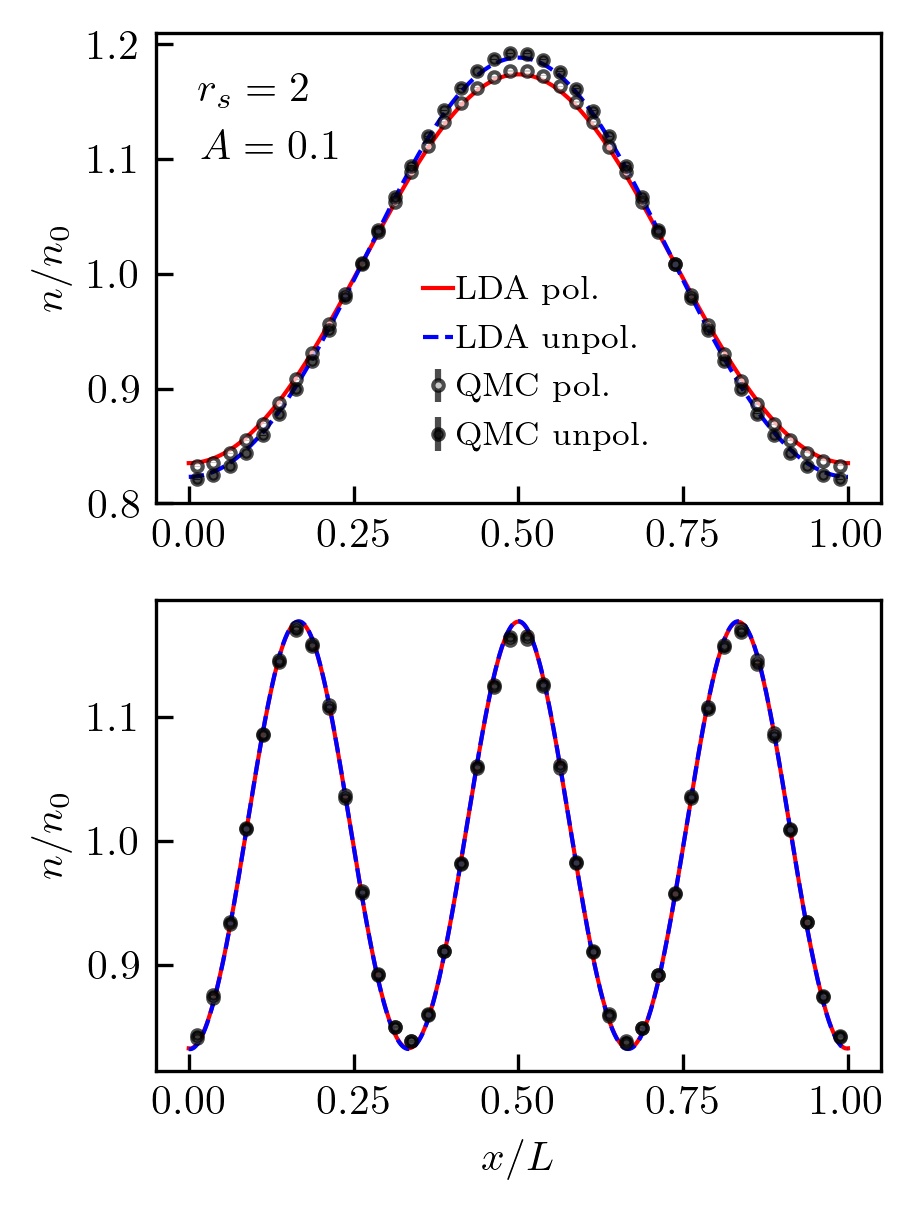}
\caption{ \label{fig:unpol} Comparing the electron density in the spin-polarized electrons gas with the spin-unpolarized case. The density is shown along the perturbation direction for $A=0.1$ and $r_s=2$ at $q_1=q_{\rm min}$ (top) and $q_1=3q_{\rm min}$ (bottom).
}
\end{figure}

While the spin-unpolarized electron gas has been assessed elsewhere~\cite{moldabekov2021relevance}, we provide a concise comparison with the spin-polarized case for completeness. 
To this end, we contrast the electron densities at $r_s=2$ and a temperature $T=T_F=12.5~{\rm eV}$ for a single harmonic perturbation with $A=0.1$ in Fig.~\ref{fig:unpol},   
The top panel corresponds to a perturbing wave number $q_1=q_{\rm min}$ and the bottom panel to $q_1=3q_{\rm min}$.
From the top panel ($q_1=q_{\rm min}$) we infer that the density perturbation induced in the spin-polarized electron gas is weaker than that in the spin-unpolarized case. 
This can be understood by recalling that the spin-polarized state has a larger Fermi energy.
Therefore, the external field has a smaller amplitude on the scale of the Fermi energy which sets the energy scale for the system. 
Contrarily at $q_1=3q_{\rm min}$ (bottom panel), the induced density perturbation is virtually identical for both the spin-polarized and spin-unpolarized cases. 
At large wave numbers, the response is dominated by single-particle effects. Consequently, both spin polarizations of the electron gas exhibit a similar response to an external field. For completeness, we note that the external field considered here does not interact with electronic spin.

\section{Computational scaling}\label{app:computational_scaling} 
\begin{figure}
\center
\includegraphics{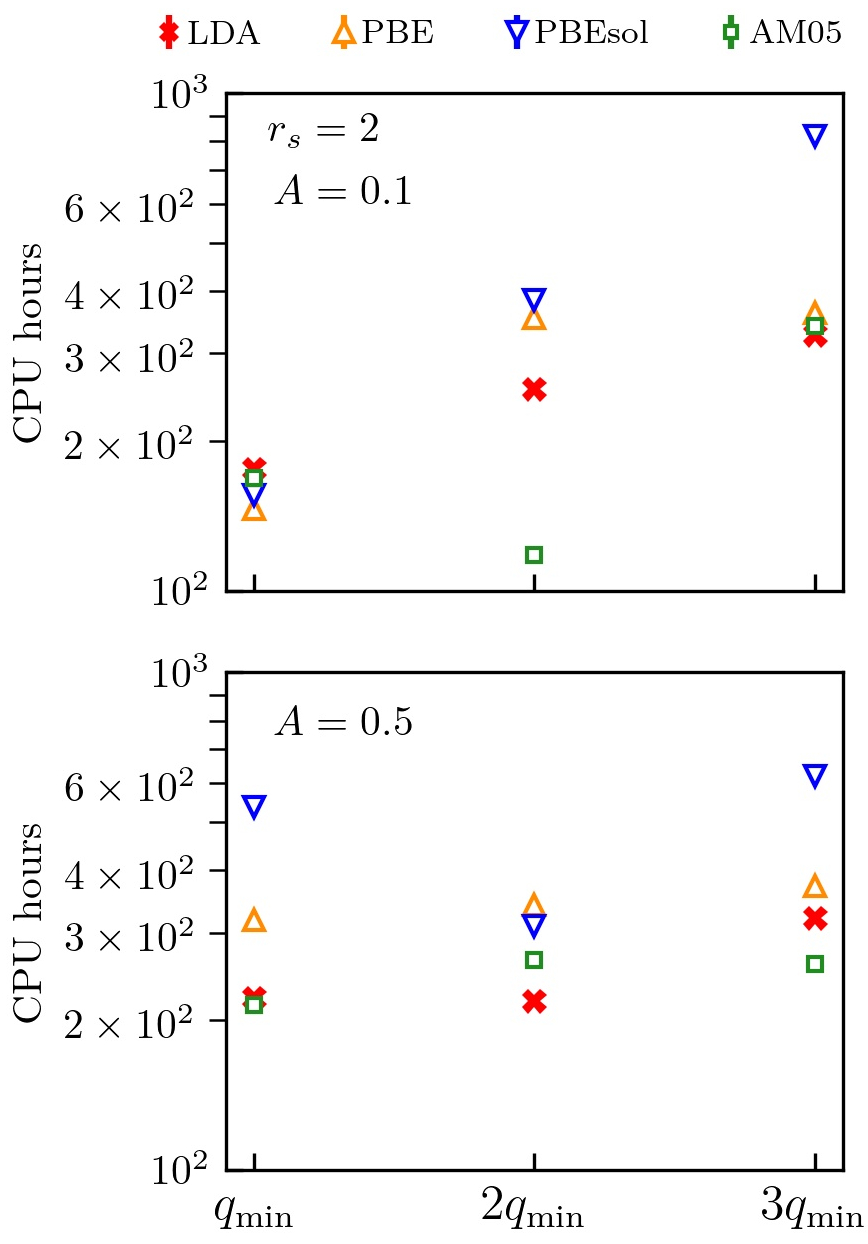}
\caption{ \label{fig:CPUrs2} Comparison of CPU hours required to obtain converged results using different XC functionals at $r_s=2$ for weakly (top) and strongly (bottom) perturbed spin-polarized electrons.
}
\end{figure}

\begin{figure}
\center
\includegraphics{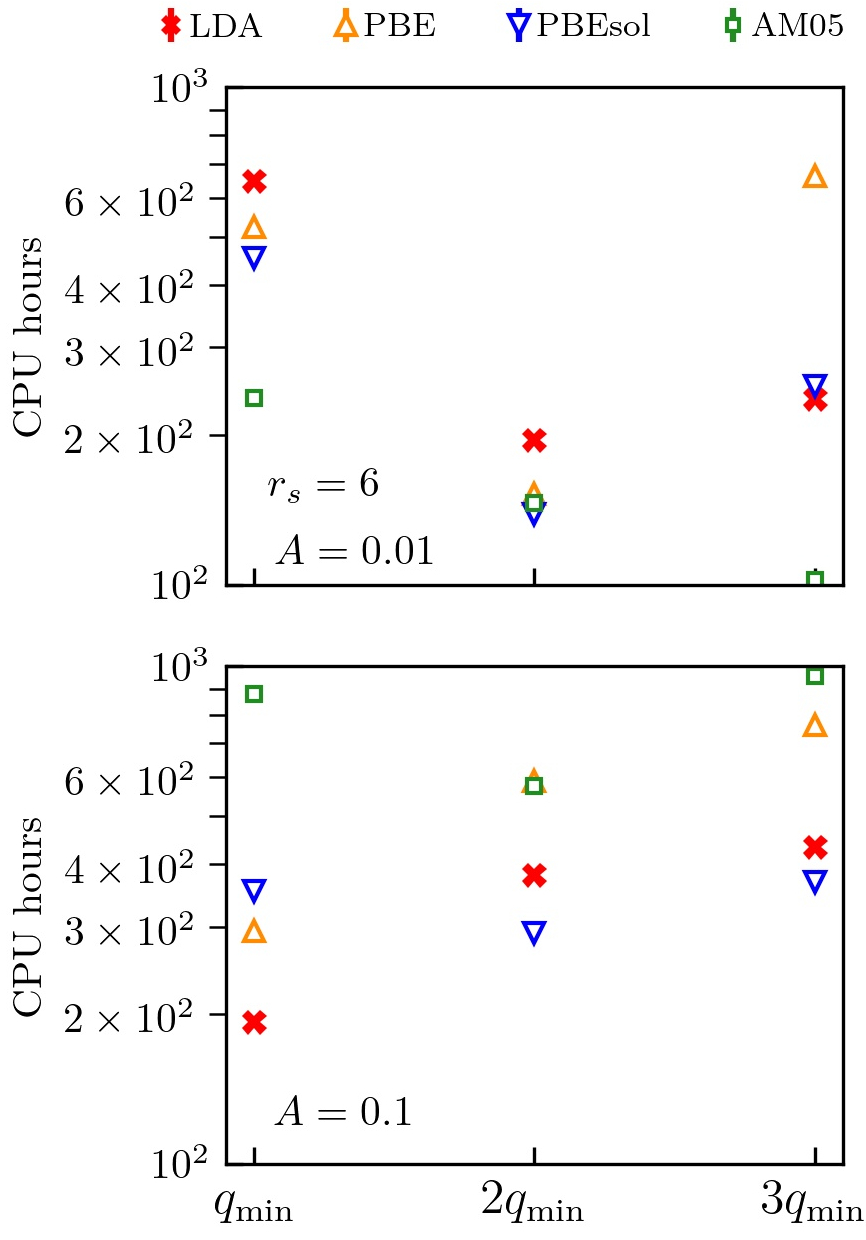}
\caption{ \label{fig:CPUrs6} Comparison of CPU hours required to obtain converged results using different XC functionals at $r_s=6$ for weakly (top) and strongly (bottom) perturbed spin-polarized electrons.
}
\end{figure}

The amount of CPU hours required to obtain converged results for different parameters at $r_s=2$ and $r_s=6$ is shown in Figs.~\ref{fig:CPUrs2} and \ref{fig:CPUrs6}.

At $r_{s}=2$, the calculations with PBEsol turned out most expensive, while the other XC functionals were similar in their computational cost (except AM05 at $A=0.1$ and $2q_{min}$). In contrast to this, at $r_s=6$, the calculations with PBEsol were overall much cheaper. Note that, when $\delta n/n_0>1$, the calculations based on AM05 required substantially more compute time than those for LDA, PBE, and PBEsol.   

\section*{Data Availability}
The data supporting the findings of this study are available on the Rossendorf Data Repository (RODARE)~\cite{data}.

\bibliography{main}

\end{document}